\documentclass[useAMS,usenatbib]{mn2e}

\usepackage{graphicx}

\title[Asphericity of Over-Luminous Type Ia Supernovae]
{Observational Characteristics and Possible Asphericity 
of Over-Luminous Type Ia Supernovae}
\author[K. Maeda and K. Iwamoto]{Keiichi Maeda$^{1}$\thanks{E-mail:
keiichi.maeda@ipmu.jp} and 
Koichi Iwamoto$^{2}$\\
$^{1}$Institute for the Physics and Mathematics of the Universe (IPMU), 
University of Tokyo, Kashiwano-ha 5-1-5, \\Kashiwa City, Chiba 277-8582, Japan\\
$^{2}$Department of Physics, College of Science and Technology, 
Nihon University, 1-8-14 Surugadai, \\Chiyoda-ku, Tokyo 101-8308, Japan}
\begin{document}

\date{}

\pagerange{\pageref{firstpage}--\pageref{lastpage}} \pubyear{2008}

\maketitle

\label{firstpage}

\begin{abstract}
A few Type Ia supernovae (SNe Ia) have been suggested to be an explosion of a super-Chandrasekhar-mass white dwarf (WD) to account for their large luminosities, requiring a large amount of $^{56}$Ni. 
However, the candidate over-luminous SNe Ia 2003fg, 2006gz, and (moderately over-luminous) SN 1991T, 
have very different observational features: 
the characteristic time-scale and velocity are very different. We examine if and how the 
diversity can be explained, by 1D {\it spherical} radiation transport calculations covering a wide range of model parameters (e.g., WD mass). The observations of SN 2006gz are naturally explained by the super-Chandrasekhar-mass model. SN 1991T represents a marginal case, which may either be 
a Chandrasekhar or a super-Chandrasekhar-mass WD explosion. 
On the other hand, the low velocity and short time-scale seen in SN 2003fg indicate that the ejecta mass is smaller than the Chandrasekhar-mass, which is an apparent contradiction to the large luminosity. We suggest that the problem is solved if the progenitor WD, and thus the SN explosion,  
is aspherical. This may reflect a rapid rotation of the progenitor star, likely a consequence of 
the super-Chandrasekhar-mass WD progenitor. 
The observed differences between SNe 2003fg and 2006gz may be attributed to different viewing 
orientations. 
\end{abstract}

\begin{keywords}
white dwarfs -- radiative transfer -- 
supernovae: individual (SN 2006gz, SN 2003fg, SN 1991T)
\end{keywords}

\section{Introduction}

Type Ia Supernovae (SNe Ia) are currently the most mature cosmological 
distance indicator 
which led to the discovery of the acceleration of the Universe 
(Riess et a. 1998; Perlmutter et al. 1999). 
Their use as distance indicators relies on the 
well calibrated light curve characteristics, 
namely a phenomenological relation between the peak luminosity and the light 
curve width ("Phillips relation" or "stretching factor"; Phillips 
1993; Perlmutter et al. 1997; Phillips et al. 1999). 

By clarifying natures of progenitor system(s) of SNe Ia 
(Livio 2000: Hillebrandt \& Niemeyer 2000; Nomoto et al. 2003), one expects to obtain 
deep knowledge on the origin of the light curve relation for the better 
luminosity calibration, as well as new application of SNe Ia to a range of 
cosmological study. 
A progenitor of normal SNe Ia (Branch et al. 1993; Li et al. 2001) 
is believed to be a white dwarf (WD) having nearly the Chandrasekhar-mass. 
Recently, a few SNe Ia have been suggested to have originated from 
a WD whose mass exceeds the Chandrasekhar-mass 
of a non-rotating WD, raising a possibility that not all 
SNe Ia are from a single class of progenitor system. 

The suggestion mainly relies on their exceptionally large luminosity,  
which requires more than $1 M_{\sun}$ of $^{56}$Ni synthesized and ejected 
during the explosion. The thermonuclear explosion converts the initial WD 
compositions partly to $^{56}$Ni. $\gamma$-rays from radioactive 
decay $^{56}$Ni $\to$ $^{56}$Co $\to$ $^{56}$Fe, through Compton scatterings 
and thermalization, power the optical emission of SNe Ia. 
Thus, the peak luminosity 
is closely related to the mass of $^{56}$Ni initially synthesized (hereafter 
$M_{\rm 56Ni}$). 

Historically, the first observationally-based suggestion for the 
super-Chandrasekhar WD explosion was made for SN Ia 1991T, as 
is summarized by Lira et al. (1998) and Fisher et al. (1999). 
However, the distance to this SN has been actively 
debated, and the smaller value is now favored (e.g., Saha et al. 2001). 
As a result, $M_{\rm 56Ni}$ required to explain the peak bolometric 
luminosity has been reduced down; 
For the reddening $E(B-V)_{\rm host} = 0.14$ and the distance modulus 
$\mu = 30.74$ as adopted by Stritzinger et al. (2006), 
the peak $V$-band magnitude is $M_V \sim -19.66$. 
Stritzinger et al. (2006) derived the mass of $^{56}$Ni, 
$M_{\rm 56Ni} \sim 0.87 M_{\odot}$, based on 
the peak bolometric luminosity $L_{\rm bol} \sim 1.7 \times 
10^{43}$ erg s$^{-1}$. 
This amount of $^{56}$Ni could be marginally explained even by 
a Chandrasekhar-mass model, although one may need an extreme, 
purely detonation-driven explosion (Khokhlov, M\"uller, \& H\"oflich 1993). 

The progress had to wait until a discovery of 
over-luminous SNe Ia, which require that $M_{\rm 56Ni} \ga 1 M_{\sun}$. 
The first candidate, SN 2003fg (SNLS-03D3bb), was reported by 
Howell et al. (2006).  
The SN reached $M_V = -19.94$ mag at peak 
and $L_{\rm bol, peak} \sim 2.6 \times 10^{43}$ erg s$^{-1}$. 
From this, Howell et al. (2006) estimated that 
$M_{\rm 56Ni} \sim 1.29 M_{\sun}$, 
using an approximate relation between the peak luminosity and $M_{\rm 56Ni}$. 
Since there should be other elements required both theoretically 
(i.e., all the well studied Chandrasekhar-mass models do not 
produce more than $1 M_{\sun}$ of $^{56}$Ni; 
Khokhlov et al. 1993; Iwamoto et al. 1999; 
R\"opke et al. 2007 and references therein) 
and observationally (i.e., the maximum-light spectrum of SN 2003fg 
was dominated by intermediate mass elements), Howell et al. (2006) argued 
that SN 2003fg should come from a super-Chandrasekhar-mass WD. 
Two other over-luminous SNe Ia, thus candidates of the super-Chandrasekhar-mass 
WD explosions, have been reported: SNe Ia 
2006gz (Hicken et al. 2007) and 2007if (Yuan et al. 2007).  
For SN 2006gz, Hicken et al. (2007) adopted $E(B-V)_{\rm host} = 0.18$ and 
$\mu = 34.95$ (but also see \S 2), deriving $M_V = -19.74$ at peak 
and $L_{\rm bol, peak} \sim 2.2 \times 10^{43}$ 
erg s$^{-1}$, and thus $M$($^{56}$Ni) $\sim 1.2 M_{\odot}$. 

In addition to the large luminosity, they are unique also in their spectra. 
Temporal sequence of the optical spectra of SN 2006gz is presented by 
Hicken et al. (2007). They identified C II $\lambda$4745, 5490, 6580, and 7324, 
in the spectra taken $10 - 14$ days before the maximum brightness, 
with the equivalent width reaching $\sim 25$\AA\ and at the velocity of $\sim 15,500$ km s$^{-1}$. 
This is the strongest evidence for unburned carbon of any SNe Ia observed so far 
(SN 1990N by Leibundgut et al. 1991; SN 2006D by Thomas et al. 2007). 
Adding to this, a Si II 6355\AA\ absorption velocity was unusually 
low well before the maximum ($\sim 12,500$ km s$^{-1}$ at $\sim 10$ 
days before the maximum). 
Without rapid decline observed in typical SNe Ia (Benetti et al. 2005), 
the Si II velocity at the maximum brightness settled down to 
$v_{\rm SiII} \sim 11,500$ km s$^{-1}$, a value typical of normal SNe Ia. 
Based on this unusually slow decline in the Si II velocity, 
Hicken et al. (2007) speculated that this might be a result of 
a dense unburned C+O envelope overlying a Si-rich region; The dense C+O 
envelope may be expected in the merger of 
two WDs leading to the formation of the super-Chandrasekhar-mass WD progenitor. 
Howell et al. (2006) also reported the probable detection of a C II feature in 
the spectrum of SN 2003fg around the maximum brightness; if pre-maximum spectra had been 
available, the C II feature may have been in the spectra early on. 
These spectroscopic features suggest that the strong C II and 
the slow evolution of Si II velocity may be a distinct feature of over-luminous 
SNe Ia. However, the spectral sequence is only available for SN 2006gz, and 
it has not been clarified to what extent these features are common in over-luminous 
SNe Ia. Furthermore, it has not been clarified yet how these features, e.g., the dense C+O 
envelope, are related to the progenitor system (e.g., see Maeda et al. 2008b for 
a caution of interpreting the C+O-rich region as an outcome of merging two WDs). 

Summarizing, the main argument for the super-Chandrasekhar model is the large 
peak luminosity and the mass of $^{56}$Ni, with spectroscopic features only indicative. 
Detailed study on observed characteristics of super-Chandrasekhar-mass 
models is missing. Clearly, the luminosity is not the only quantity that 
is dependent on the underlying models. 
The suggestion by Howell et al. (2006), about the Super-Chandrasekhar-model for 
SN 2003fg, is actually based on another observed character, i.e., the velocity 
of the expanding SN materials 
as deduced from its spectra. SN 2003fg showed exceptionally small velocity 
around the maximum brightness ($\sim 8,000$ km s$^{-1}$) compare to 
that of normal SNe Ia ($\ga 10,000$ km s$^{-1}$). At the first look, 
this seems fully consistent with the expectation, since the super-Chandrasekhar-mass 
WD provides the large binding energy, and thus the small kinetic energy per mass 
as compared to an explosion of a Chandrasekhar-mass WD. 
Jeffery, Branch, and Baron (2006) confirmed 
this statement on the relation between the WD mass and the velocity-scale. 

Although the argument may sound satisfactory, 
it is not a whole story. 
Studying emission processes should add further 
constraints on the underlying models. 
There are two over-luminous SNe Ia whose observed characteristics are 
available in the literature, i.e., 2003fg and 2006gz. 
Extensive observational data set for the moderately over-luminous 
SN 1991T is also available.  
By examining these 
observations (\S 2), we clarify that 
they have quite different properties. For example, SN 2006gz showed 
the velocity similar to normal SNe Ia - then, the argument for the 
super-Chandrasekhar-mass model for SN 2006gz may be 
flawed.\footnote{Indeed, it is shown in this paper that the low 
velocity seen in SN 2003fg has an apparent 
problem in the super-Chandrasekhar-mass model.} 
The aim of this paper is to examine if the observed characteristics can 
be explained by super-Chandrasekhar-mass models, by means of radiation 
transfer calculations. We especially focus on SNe 2003fg and 2006gz. 
In \S 3, we describe SN Ia models 
and summarize a method for the radiation transfer calculations.  
Results based on an extensive set of model calculations are shown in \S 4. 
Discussion is given in \S 5, where 
we examine uncertainties involved in our calculations. 
We also discuss the nature 
of these peculiar over-luminous SNe Ia. The emphasis is placed on our 
finding that not every observed characteristics can be interpreted 
within the context of the simple (spherical) super-Chandrasekhar-mass WD models, 
and on our suggestion 
that the progenitor should largely deviate from spherical symmetry to account for 
the observed features 
(at least for SN 2003fg). Implications on SN 1991T are also discussed. 
The paper is closed in \S 6 with concluding remarks.

\section{OBSERVATIONAL CHARACTERISICS}

In this section, we summarize observed features relevant to this work, 
for SNe 1991T, 2003fg, and 2006gz. 
Howell et al. (2006) presented their observational data 
of SN Ia 2003fg. The multi-band light 
curves are fitted well by $k$-corrected template light curves 
with the stretching factor (Perlmutter et al. 1997, 1999) of $s = 1.13$. 
The corresponding $\Delta m_{15}^{B}$ (the magnitude change in 
the first 15 days past $B$ maximum: Phillips 1993) is the following: 
$\Delta m_{15}^{B} = 0.84 \pm 0.02$, using equation (5) of 
Perlmutter et al. (1997)\footnote{Hicken et al. (2007) 
derived $\Delta m_{15}^{B} \sim 0.9$ for SN 2003fg in their Figure 4. 
Although this is different from what we derived here, 
the expected $t_{+1/2}$ is $\sim 13$ days, which is well within the errors 
of our estimate}. 
We then convert $\Delta m_{15}^{B}$ to $t_{+1/2}$, the time since the 
maximum luminosity to half the maximum luminosity as measured in a 
bolometric light curve (Contardo, Leibundgut, \& Vacca 2000). This is possible 
thanks to a correlation between these two quantities.  
Note that $t_{+1/2}$ in this paper is a {\it post-maximum} 
quantity, not the {\it rise time} frequently used in the related field. 
For the conversion, we fit an observationally derived 
set of ($\Delta m_{15}^{B}$, $t_{+1/2}$) for nearby 9 
SNe Ia presented in 
Contardo et al. (2000), assuming a functional 
form of $t_{+1/2} = a \Delta m_{15}^{B} + b$. 
We obtain $a = -4.52 \pm 0.786$ and $b = 17.3 \pm 1.03$.  
Using this relation, $t_{+1/2} = 13.5 \pm 1.7$ days for SN 2003fg. 
The derivation of $t_{+1/2}$ in SN Ia 2006gz is straightforward. 
The {\it UBVRI} bolometric curve is presented by Hicken et al. 
(2007)\footnote{The data file  is presented at 
http://www.cfa.harvard.edu/supernova/SNarchive.html .}, 
and it directly gives $t_{+1/2} = 18$ days. 
The light curve width is very different between the two SNe Ia. 

They are very different in characteristic velocity, too. 
SN 2003fg showed a Si II absorption velocity ($v_{\rm Si II}$) of 
$8,000 \pm 500$ km s$^{-1}$ 
around maximum brightness (Howell et al. 2006). 
The same value for 
SN 2006gz is in the range $11,000 - 12,000$ km s$^{-1}$ (Hicken et al. 2007). 

Typically, normal SNe Ia have $t_{+1/2}$ in the range of $\sim 9 - 14$ days 
(e.g., Contardo, et al. 2000)
and $v_{\rm Si II}$ in $\sim 10,000 - 14,000$ km s$^{-1}$ 
(e.g., Hachinger, Mazzali, \& Benetti 2006).  
If we focus on SNe Ia with $M_{\rm 56Ni} \sim 0.6 M_{\sun}$, 
the Phillips relation indicates that $\Delta m_{15}^{B} \sim 1.1$ 
(Stritzinger et al. 2006). 
This corresponds to $t_{+1/2} = 12.1 \pm 1.8$ using the relation we derived 
above. The examples are SNe 2003du and 1990N (e.g., Stritzinger et al. 2006), 
both having $v_{\rm Si II} \sim 10,000 - 11,000$ km s$^{-1}$. 

The time scale, $t_{+1/2}$, in SN 2003fg is at about the upper bound of normal 
SNe Ia, while $v_{\rm Si II}$ is much lower. $t_{+1/2}$ in SN 2006gz is 
well above the range seen in normal SNe Ia, while $v_{\rm Si II}$ is 
similar to normal cases. 
As for the bolometric magnitude at the maximum light ($L_{\rm bol, peak}$), 
SN 2003fg has been claimed to be over-luminous; 
$L_{\rm bol, peak} \sim (2.5 - 2.8) \times 10^{43}$ erg s$^{-1}$ 
(Howell et al. 2006). 
SN 2006gz has also been claimed to be over-luminous; 
$L_{\rm bol, peak} = (1.8 - 2.6) \times 10^{43}$ erg s$^{-1}$ 
assuming $E(B-V)_{\rm host} = 0.18$ and $\mu = 34.95$  (Hicken et al. 2007). 
However, there is a caveat for $L_{\rm bol, peak}$ in SN 2006gz; 
Maeda et al. (2008b) pointed out that the peak luminosity of SN 2006gz 
derived by Hicken et al. (2007) involves large uncertainty, because  
the host-galaxy extinction has not been well constrained. 
If the host extinction would be totally negligible, then 
$M_{V} \sim -19.2$ at peak and $L_{\rm bol, peak} \sim 1.3 \times 
10^{43}$ erg s$^{-1}$, which is relatively bright but not 
over-luminous\footnote{This uncertainty highlights the importance to investigate the 
radiation process and to obtain constraints independent from the peak luminosity, 
as is aimed in this paper.}. 

For SN 1991T, the following values are relevant. 
$t_{+1/2} = 14.0$ given by Contardo et al. (2000), based on 
compilation of multi-band light curve. 
$v_{\rm Si II} \sim 10,000 - 11,000$ km s$^{-1}$ 
(Hachinger et al. 2006 and references therein). 
In these properties, SN 1991T represents 
an intermediate case between SNe 2003fg and 2006gz, 
and indeed close to those of normal SNe Ia. 
The peak luminosity with the favored distance modulus 
is $L_{\rm bol, peak} \sim 1.7 \times 10^{43}$ erg s$^{-1}$ (Stritzinger et al. 2006) 
which is at the upper bound of normal SNe Ia, although the original 
suggestion for the super-Chandrasekhar-mass model was based on the 
brighter estimate ($L_{\rm bol, peak} \sim 2.3 \times 10^{43}$ erg s$^{-1}$; 
Fisher et al. 1999; Contardo et al. 2000).

Summarizing, the two most probable candidate super-Chandrasekhar-mass 
WD explosions, i.e., SNe 2003fg and 2006gz, have very different observational 
characteristics, although this fact has not been emphasized in the literature.

\section{Method and Models}

\subsection{SN Ia Models}
We construct SN Ia models starting with the density structure of 
the W7 model (Nomoto, Thielemann, \& Yokoi 1984) which reproduces basic observational 
features of normal SNe Ia (Branch et al. 1985). 
Assuming the homologous expansion which should be a good approximation 
for an explosion of a compact progenitor, 
the density distribution as a function of velocity is uniquely 
determined from the normalized reference density distribution (i.e., W7), by 
specifying the following model parameters: 
\begin{enumerate}
\item $M_{\rm wd}$: the mass of the WD. 
\item $\rho_{\rm c}$: the WD central density at the burning ignition. 
\item $f_{\rm ECE}$: the mass fraction of Fe-peak elements 
produced by strong electron captures, i.e., Fe-peak elements 
excluding $^{56}$Ni (e.g., $^{58}$Ni, $^{56}$Fe, $^{54}$Fe). 
\item $f_{\rm 56Ni}$: the mass fraction of $^{56}$Ni. 
\item $f_{\rm IME}$: the mass fraction of partially burned intermediate mass 
elements like Mg, Si, and S. 
\end{enumerate}
The mass fraction of unburned C+O materials ($f_{\rm CO}$) is then simply 
given by $f_{\rm CO} = 1 - f_{\rm ECE} - f_{\rm 56Ni} - f_{\rm IME}$. 

The procedure of the model construction follows that presented in 
Jeffery et al. (2006) (see also Howell et al. 2006 and Maeda et al. 2008b). 
For given $M_{\rm wd}$ and $\rho_{\rm c}$, we compute the binding 
energy of the WD ($E_{\rm b}$) by the formulae given by 
Yoon \& Langer (2005), who examined a sequence of 
structure of a rotating WD. The energy produced by the 
nuclear burning ($E_{\rm nuc}$) is given by $M_{\rm wd}$, $f_{\rm ECE}$, $f_{\rm 56Ni}$, 
and $f_{\rm IME}$, by a simple relation 
\begin{equation}
E_{\rm nuc}/(10^{51} {\rm erg}) = (1.74 f_{\rm ECE} + 1.56 f_{\rm 56Ni} + 1.24 f_{\rm IME}) 
M_{\rm wd}/M_{\sun} \ .
\end{equation}
The kinetic energy is then simply 
\begin{equation}
E_{\rm K} = E_{\rm nuc} - E_{\rm b} \ .
\end{equation}
The reference density structure is then scaled in a self-similar manner, i.e., 
\begin{equation}
\rho (v) \propto M_{\rm wd}^{5/2} E_{\rm K}^{-3/2} \ , 
\end{equation}
and 
\begin{equation}
v \propto M_{\rm wd}^{-1/2} E_{\rm K}^{1/2} \ , 
\end{equation}
where $v$ is a velocity of a Lagrangian fluid element, and $\rho (v)$ the 
density there. 

Now that we have specified the density structure, 
we are left to specify the distribution of 
elements. We examine three extreme cases so that our models should cover 
the real situation anyway. 
\begin{enumerate}
\item Stratified (model sequence "a"): 
characteristic burning layers are totally separated and stratified. 
The electron capture region (where the mass fraction of stable Fe-peak elements 
is set to be unity) is at the centre, surrounded by the $^{56}$Ni-rich region 
(the mass fraction of $^{56}$Ni is set to be unity), then by the partially burned 
layer (mass fractions of Si and S are set to be 0.7 and 0.3, respectively), 
and by the unburned C+O layer at the outermost region (mass fractions of 
C and O are set to be 0.5 and 0.5, respectively). 
\item Mixing in the Fe-rich region (model sequence "b"): 
Basically the same with the model sequence "a", except that the innermost 
electron capture region and the $^{56}$Ni region are assumed to be fully mixed 
in the composition structure. 
\item Fully mixing (model sequence "c"): 
The compositions are fully and homogeneously mixed throughout the ejecta. 
\end{enumerate}
The models "a" and "b" are most likely the case at least in normal SNe Ia, 
in a viewpoint of observed spectral evolution 
(Stehle et al. 2005; Mazzali et al. 2008). There are observational hints 
favoring the models "a" than "b" for some SNe Ia (H\"oflich et al. 2004; 
Motohara et al. 2006). 
Examples of the density and composition structures are shown in 
Figure 1. 

\begin{figure}
\includegraphics[width=84mm]{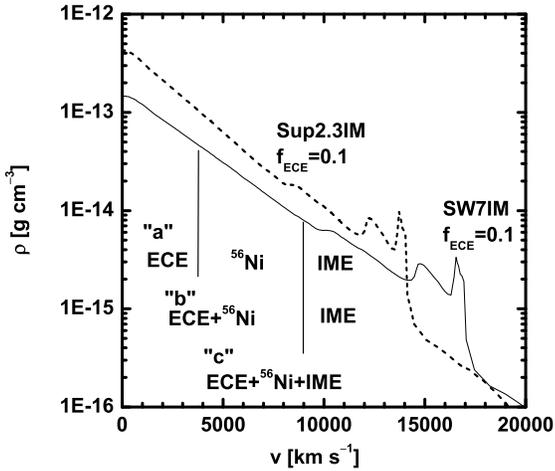}
\caption{Examples of the density structure are 
shown for SW7IM (with $f_{\rm ECE} = 0.1$; solid) and for Sup2.3IM 
(with $f_{\rm ECE} = 0.1$; dashed). Also shown is the composition structure 
of SW7IM, for the three different mixing prescription.}
\label{fig1}
\end{figure}

We have examined an extensive set of models as shown in Table 1. 
$M_{\rm wd}$ is varied from $1.39 M_{\sun}$ to $2.6 M_{\sun}$ separated by 
$0.3 M_{\sun}$ each. 
$\rho_{\rm c}$ is set to be $3 \times 10^{9}$ g cm$^{-3}$ in most models, as 
the model sequence with varying $\rho_{\rm c}$ can be self-similar (only 
different in $E_{\rm K}$). The dependence of the observed behaviors 
on $\rho_{\rm c}$ is examined only for the models with $M_{\rm wd} = 2 M_{\sun}$, 
and can be generalized to different masses in a straightforward way. 
$f_{\rm 56Ni}$ is fixed for given $M_{\rm wd}$, corresponding to 
$M_{\rm 56Ni} = 0.6 M_{\sun}$ for the Chandrasekhar-mass models ($M_{\rm wd} = 1.39 M_{\sun}$) 
and $M_{\rm 56Ni} = 1.0 M_{\sun}$ for the super Chandrasekhar-mass models ($M_{\sun} > 1.39 M_{\sun}$). 
$f_{\rm ECE}$ is varied from 0 (or 0.1 so that $E_{\rm K}$ should be positive) 
to $1.0 - f_{\rm 56Ni}$, corresponding to the complete burning of whole ejecta 
to the Fe-peak elements, separated by 0.1 each. 
Either $f_{\rm IME}$ or $f_{\rm CO}$ is set to be zero to reduce the number of model 
calculations. These two represent the two extreme cases, and thus should cover the 
real situation. The total number of models we construct here is then 246, 
including models with different $f_{\rm ECE}$ and different mixing prescription.

\begin{table*}
 \caption{SN Ia Model.}
 \label{tab:model}
 \begin{tabular}{@{}lllllllll@{}}
 \hline
Name & $M_{\rm wd}$ & $\rho_{c}$ &
$f_{\rm ECE}$ & $f_{\rm 56Ni}$ &
$f_{\rm IME}$ & $f_{\rm CO}$ & $M_{\rm 56Ni}$ &
$E_{\rm k}$ \\
 \hline
SW7IM  & 1.39 & 3.0e9 & (0 - 0.57) & 0.43 & (0.57 - 0) & 0 & 0.6 & 1.40 - 1.80 \\
LW7IM  & 1.39 & 3.0e9 & (0 - 0.28) & 0.72 & (0.28 - 0) & 0 & 1.0 & 1.53 - 1.73 \\
Sup1.7IM  & 1.70 & 3.0e9 & (0 - 0.41) & 0.59 & (0.41 - 0) & 0 & 1.0 & 1.55 - 1.90 \\
Sup2IM  & 2.00 & 3.0e9 & (0 - 0.50) & 0.50 & (0.50 - 0) & 0 & 1.0 & 1.56 - 2.06 \\
Sup2IM10  & 2.00 & 1.0e10 & (0 - 0.50) & 0.50 & (0.50 - 0) & 0 & 1.0 & 1.24 - 1.74 \\
Sup2.3IM  & 2.30 & 3.0e9 & (0 - 0.57) & 0.43 & (0.57 - 0) & 0 & 1.0 & 1.56 - 2.21 \\
Sup2.6IM  & 2.60 & 3.0e9 & (0 - 0.62) & 0.38 & (0.62 - 0) & 0 & 1.0 & 1.56 - 2.36 \\
SW7CO  & 1.39 & 3.0e9 & (0 - 0.57) & 0.43 & 0 & (0.57 - 0) & 0.6 & 0.42 - 1.80 \\
LW7CO  & 1.39 & 3.0e9 & (0 - 0.28) & 0.72 & 0 & (0.28 - 0) & 1.0 & 1.05 - 1.73 \\
Sup1.7CO  & 1.70 & 3.0e9 & (0 - 0.41) & 0.59 & 0 & (0.41 - 0) & 1.0 & 0.69 - 1.90 \\
Sup2CO  & 2.00 & 3.0e9 & (0 - 0.50) & 0.50 & 0 & (0.50 - 0) & 1.0 & 0.32 - 2.06 \\
Sup2CO10  & 2.00 & 1.0e10 & (0 - 0.50) & 0.50 & 0 & (0.50 - 0) & 1.0 & 0.001 - 1.74 \\
Sup2.3CO  & 2.30 & 3.0e9 & (0.10 - 0.57) & 0.43 & 0 & (0.47 - 0) & 1.0 & 0.33 - 2.21 \\
Sup2.6CO  & 2.60 & 3.0e9 & (0.10 - 0.62) & 0.38 & 0 & (0.52 - 0) & 1.0 & 0.01 - 2.36 \\
 \hline
\end{tabular}

\medskip 
Models with the W7 reference density distribution. 
For the description of the parameters, see the main text. 
Each column represents a set of models: 
For example, model SW7IM is examined with three 
different mixing prescription (mixing "a", "b", and "c"; see the main text). 
For each mixing, a set of ($f_{\rm ECE}$, $f_{IME}$) -- (0, 0.57), (0.1, 0.47), 
(0.2, 0.37), (0.3, 0.27), (0.4, 0.17), (0.5, 0.07), and (0.57, 0) -- are examined. 

\end{table*}

\subsection{Radiation Transfer}
For each model which is descritized in 172 radial zones, 
we performed radiation transfer 
calculations based on a numerical code presented 
in Iwamoto et al. (2000).  
The important assumption in our calculations 
is spherical symmetry. The transport of $\gamma$-rays 
from radioactive $^{56}$Ni/Co/Fe decay is solved under a gray 
atmosphere approximation with $\kappa_{\gamma} = 0.027$ cm$^{2}$ g$^{-1}$, 
which should be a sufficiently accurate approximation for our present purpose 
(Maeda 2006a). With the energy input by $\gamma$-rays, the optical photon 
transport is solved again under the gray approximation. 
Opacity is provided by Thomson scatterings and line scatterings, i.e., 
\begin{equation}
\kappa = \kappa_{e-} + \kappa_{\rm line} \ .
\end{equation}
The ionization is determined by solving the Saha equation (i.e., in LTE), then 
used to determine $\kappa_{e-}$ at each time step and in each radial grid. 

The prescription of $\kappa_{\rm line}$ is more complicated and uncertain, 
as it is affected by many weak lines whose atomic data are experimentally not 
well known (Woosley et al. 2007). 
In this paper, we follow a simplified, phenomenological 
prescription based on modeling observations of normal SNe Ia, and restrict our selves to 
focus on bolometric light curves.  
The main advantage of this treatment is that the bolometric light curve 
is insensitive to a color evolution which can be quite complicated. 
We follow Mazzali et al. (2001): 
\begin{equation}
\kappa_{\rm line} = 0.5 \times 
\left[0.25 (X_{\rm ECE}+X_{\rm 56Ni}) + 0.025 (X_{\rm IME} + X_{\rm CO})\right] 
\ {\rm cm}^{2} {\rm g}^{-1} \ .
\end{equation}
Here, $X_{i}$ denotes a local (position-dependent) mass fraction. 
The coefficient (0.5) is added to the original prescription of Mazzali et al. (2001) 
so that the SW7 model sequence satisfies the observational constraint of 
normal SNe Ia with $M_{\rm 56Ni} \sim 0.6 M_{\sun}$ (\S 4).

\section{Result}

\subsection{Fading Rate and Photospheric Velocity} 

Figure 2 shows examples of the synthetic bolometric light curves. 
Shown in the figure are the models with the mixing "a" (stratified) 
and $f_{\rm ECE} = 0.1$. If the other parameters are similar, 
models with larger $M_{\rm wd}$ have broader light curve shape 
around the maximum brightness, i.e., $t_{+1/2}$ is larger 
for larger $M_{\rm wd}$. Also, the photospheric velocity at the 
maximum luminosity ($v_{\rm peak}$) is smaller for 
larger $M_{\rm wd}$ because of smaller $E_{\rm K}/M_{\rm wd}$ 
(note that the velocity in Sup1.7IM is larger than SW7IM because of the 
larger amount of $M_{\rm 56Ni}$ and thus larger $E_{\rm K}/M_{\rm wd}$). 

\begin{figure}
\includegraphics[width=84mm]{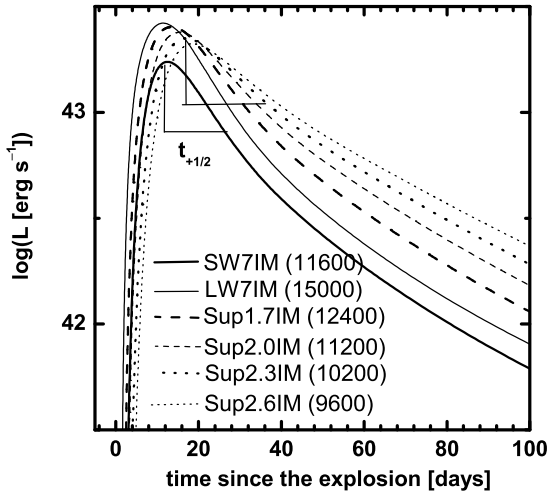}
\caption{Examples of the synthesized bolometric light curves. 
Shown here are the models with the mixing "a" (stratified) and 
$f_{\rm ECE} = 0.1$; SW7IM (thick-solid), LW7IM (thin-solid), 
Sup1.7IM (thick-dashed), Sup2.0IM (thin-dashed), 
Sup2.3IM (thick-dotted), and Sup2.6IM (thin-dotted). Shown in 
parentheses are the photospheric velocity at the maximum bolometric 
brightness ($v_{\rm peak}$ km s$^{-1}$). }
\label{fig2}
\end{figure}

This example shows that the relation between 
$t_{+1/2}$ and $v_{\rm peak}$ can provide a strong tool to check 
the validity of a model as compared with the observations, although 
this has not been examined for SNe 2003fg and 2006gz in the previous studies. 
The constraints are the following: (1) A model should reproduce $t_{+1/2}$ derived in 
\S 2. (2) The photospheric velocity at the maximum luminosity in a model 
($v_{\rm peak}$) should be {\it equal to or less than} observed $v_{\rm Si II}$, since 
the Si II absorption must be formed above the photosphere. 

Figure 3 shows ($t_{+1/2}$, $v_{\rm peak}$) for all the models, 
except for Sup2.0IM10 and Sup2.0CO10\footnote{We have found that 
($t_{+1/2}$, $v_{\rm peak}$) of Sup2.0IM10/Sup2.0CO10 
can be obtained by simply shifting those of Sup2.0IM/CO, along the curve defined by 
the mixing "a" to the bottom-right direction. This is because 
larger $\rho_{\rm c}$ is equivalent to lower $E_{\rm K}$. Thus changing $\rho_{\rm c}$ 
can be mimicked by changing $E_{\rm K}$ (i.e., by changing $f_{\rm ECE}$).}. 
The SW7IM/CO model sequence (with $M_{\rm 56Ni} = 0.6 M_{\sun}$) approximately satisfies the 
($t_{+1/2}$, $v_{\rm Si II}$) constraint of the normal SNe Ia with $M_{\rm 56Ni} 
\sim 0.6 M_{\sun}$ (e.g., SNe 2003du, 1990N, as the observed range shown in Figure 3). 

\begin{figure}
\includegraphics[width=84mm]{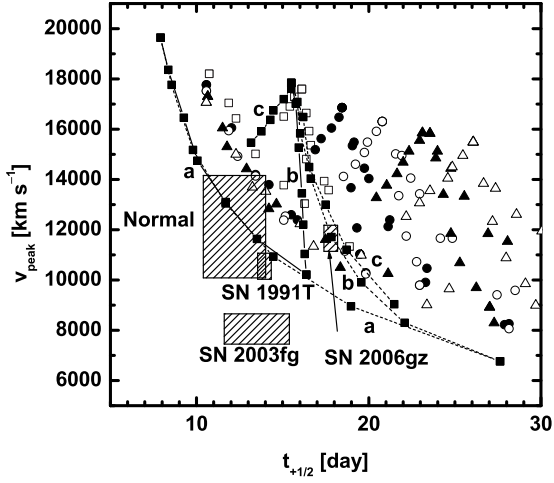}
\caption{($t_{+1/2}$, $v_{\rm peak}$) shown for the models. 
The meaning of the symbols are the following: SW7IM and SW7CO (filled squares), 
LW7IM and LW7CO (open squares), Sup1.7IM and Sup1.7CO (filled circles), 
Sup2.0IM and Sup2.0CO (open circles), Sup2.3IM and Sup2.3CO (filled triangles), 
Sup2.6IM and Sup2.6CO (open triangles). The SW7 models with different 
$f_{\rm ECE}$ are connected by the lines, separately for 
different mixing prescription 
("a", "b", and "c"); solid lines for SW7IM and dotted lines for SW7CO. 
The observed range for normal SNe Ia with $M_{\rm 56Ni} \sim 0.6 M_{\sun}$ 
is shown by the shaded region. The observed characteristics of SNe 2003fg, 2006gz,  
and 1991T, are also plotted as the shaded regions with observed uncertainty evaluated in \S 2. 
}
\label{fig3}
\end{figure}

Important discovery in Figure 3 is that 
even a single model does not satisfy the observed characteristics of SN 2003fg. 
On the other hand, SN 2006gz can be explained by many models 
(note that observed $v_{\rm Si II}$ can be larger than synthetic $v_{\rm peak}$), 
and most naturally by 
Super-Chandrasekhar models having $M_{\rm 56Ni} \sim 1 M_{\sun}$, which results 
in $v_{\rm peak} \sim v_{\rm Si II}$ for the favored non-mixing case ("a"). 
We have examined an extensive set of models including extreme cases, so that 
our failure to reproduce the observed characteristics of SN 2003fg represents 
a principal difficulty in the super-Chandrasekhar model for this SN Ia. 

We found that the lower boundary in the $t_{+1/2} - v_{\rm peak}$ plot, covered by 
varying various parameters for given $M_{\rm wd}$, 
is represented by a model sequence of "a" with only varying $f_{\rm ECE}$. 
In the mixing cases ("b" and "c"), $^{56}$Ni is mixed down to the low velocity, 
and thus the diffusion time-scale becomes long as compared with the stratified case "a". 
This results in larger $t_{+1/2}$ in mixing "b", "c" than "a". 
We also found that the curves, obtained by varying $f_{\rm ECE}$, 
almost overlap between the cases "IM" ($f_{\rm CO} = 0$) and "CO" ($f_{\rm IME} = 0$) 
in the $t_{+1/2} - v_{\rm peak}$ plane, and the ones with $f_{IME} = 0$ 
(e.g., SW7CO) cover the wider range of parameter space. 

Thus, the model sequence with the mixing "a" and $f_{IME} = 0$ form the lower 
boundary as we vary the value for $f_{\rm ECE}$, in the $t_{+1/2} - v_{\rm peak}$ 
plane for given $M_{\rm wd}$ and $M_{\rm 56Ni}$. If the observationally derived 
set of ($t_{+1/2}$, $v_{\rm Si II}$) is above the curve, the model sequence 
is acceptable (i.e., there are observationally acceptable combinations of parameters 
for given $M_{\rm wd}$ and $M_{\rm 56Ni}$). If it is below the curve, the combination 
of $M_{\rm wd}$ and $M_{\rm 56Ni}$ should be rejected. 
Figure 4 shows the lower boundary of $v_{\rm peak}$ as a function of $t_{+1/2}$ for 
different $M_{\rm wd}$ (note that $M_{\rm 56Ni}$ is fixed for given $M_{\rm wd}$ in 
this paper). 

\begin{figure}
\includegraphics[width=84mm]{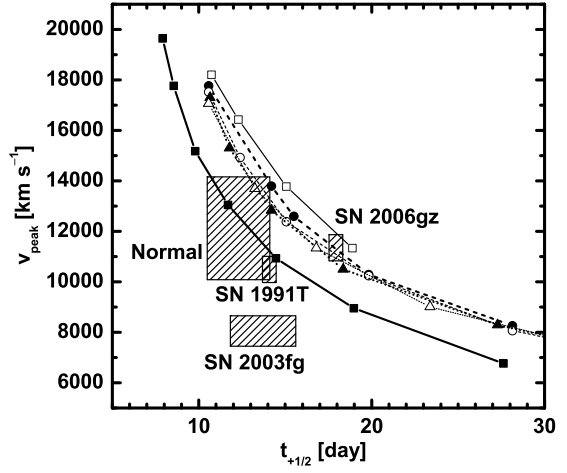}
\caption{The lower boundary in the $t_{+1/2} - v_{\rm peak}$ plane 
shown for the W7-sequence models with various $M_{\rm wd}$. 
The meaning of the symbols are the following: SW7CO (filled squares and thick solid line), 
LW7CO (open squares and thin solid line), 
Sup1.7CO (filled circles and thick dashed line), 
Sup2.0CO (open circles and thin dashed line), 
Sup2.3CO (filled triangles and thick dotted line), 
Sup2.6CO (open triangles and thin dotted line). 	
For the observed values (shaded regions), see the caption of Figure 3.  
}
\label{fig4}
\end{figure}

Figure 4 shows that the curve is similar for models having the same $M_{\rm 56Ni}$. 
The behavior can be explained as follows to the first approximation. 
The time-scale of the light curve evolution around peak is scaled 
as follows (Arnett 1982). 
\begin{equation}
t_{+1/2} \propto \kappa^{1/2} M_{\rm wd}^{3/4} E_{\rm K}^{-1/4} \ , 
\end{equation}
where $\kappa$ is the opacity averaged over the ejecta. 
The estimate of the photospheric velocity at the maximum brightness 
is complicated. For the present demonstration purpose, it is 
enough to assume that this is scaled with the average velocity, i.e., 
\begin{equation}
v_{\rm peak} \propto M_{\rm wd}^{-1/2} E_{\rm K}^{1/2} \ .
\end{equation}
Then, combining these two expressions, we obtain the following relation: 
\begin{equation}
v_{\rm peak} \propto \kappa M_{\rm wd} t_{+1/2}^{-2} \ .
\end{equation}
The opacity is provided by Fe-peak elements, and 
stable Fe-peak elements below the $^{56}$Ni-rich region does not significantly 
contribute to the opacity in non-mixing case "a". Thus, 
\begin{equation}
v_{\rm peak} \propto M_{\rm 56Ni} t_{+1/2}^{-2} \ . 
\end{equation}
Therefore, the lower boundary (i.e., non-mixing case "a") 
in the $t_{+1/2} - v_{\rm peak}$ plane for a given model 
is basically determined by $M_{\rm 56Ni}$. Note that the curves representing the lower 
boundary in Figure 4 do not exactly behaves as $v_{\rm peak} \propto t_{+1/2}^{-2}$, 
because the models with different 
$f_{\rm ECE}$ are not self-similar to one another. Specifically, a model with lower $f_{\rm ECE}$ 
(thus larger $t_{+1/2}$ and lower $v_{\rm peak}$) has $t_{+1/2}$ larger than expected by this simple 
analytic relation, since they have more centrally concentrated $^{56}$Ni distribution leading to 
the larger diffusion time-scale. 
 
One may wonder if our results are consistent with observations of normal SNe Ia and 
some peculiar SNe Ia. First of all, we emphasize that we restrict ourselves 
to investigate Chandrasekhar models with $M_{\rm 56Ni} \ga 0.6 M_{\odot}$ 
and super-Chandrasekhar models with $M_{\rm 56Ni} \sim 1 M_{\odot}$. 
As such, our calculations do not directly cover (a) detailed variation in normal SNe Ia, 
and especially (b) sub-luminous SNe Ia which requires $M_{\rm 56Ni} < 0.6 M_{\odot}$. 

However, these issues are consistent with our results, and thus it is safe to use 
our calculations to model the super-Chandrasekhar SNe Ia candidates: 
(a) A range of ${t_{+1/2}}$ seen in normal SNe Ia can be explained 
by varying $M_{\rm 56Ni}$. For example, if we take 
$M_{\rm 56Ni} < 0.6 M_{\sun}$, the model sequence moves to the left 
[i.e., to smaller $t_{+1/2}$; equation (10)]. 
This also reduces the peak luminosity as $^{56}$Ni is the energy source.  
In this way, ($t_{+1/2}$, $v_{\rm Si II}$) of normal SNe Ia can basically be explained, 
simultaneously satisfying the Phillips relation (Mazzali et al. 2001).
(b) There are two SNe Ia showing $v_{\rm peak}$ smaller than SN 2003fg. 
They are "sub-luminous" peculiar SNe Ia 2002cx (Li et al. 2003; Branch et al. 2004; 
Jha et al. 2006) 
and 2005hk (Jha et al. 2006; Phillips et al. 2007; Sahu et al. 2008), 
suggested to be a less energetic explosion 
with smaller mass of $^{56}$Ni than normal. 
SN 2005hk showed $v_{\rm SiII} \sim 6,000$ km s$^{-1}$ 
($\sim v_{\rm peak}$ according to the spectrum modeling; Sahu et al. 2008) and 
$t_{+1/2} \sim 20$ days. 
Some of our models have small $E_{\rm K}$ 
(Tab. 1), which results in small $v_{\rm peak}$ and large $t_{+1/2}$. For example, 
Figure 4 shows that for sufficiently small $E_{\rm K}$, $v_{\rm peak} \sim 8,000$ 
km s$^{-1}$ {\it and} $t_{+1/2} \sim 30$ days for $M_{\rm 56Ni} \sim 1 M_{\sun}$. 
It has been estimated that $M_{\rm 56Ni} \sim 0.2 M_{\sun}$ 
in these faint SNe Ia (Phillips et al. 2007; Sahu et al. 2008), 
and a model with such a small amount of $^{56}$Ni should be considered. 
From equation (10), we see that $t_{+1/2} \propto M_{\rm 56Ni}^{1/2} v_{\rm peak}^{-1/2}$. 
Applying our result to this relation, 
we expect that $t_{+1/2} \sim 16 - 20$ days for $M_{\rm 56Ni} \sim 0.2 - 0.3 M_{\sun}$ 
and for $v_{\rm peak} \sim 6,000$ km s$^{-1}$, 
as is consistent with the observational features of SN 2005hk. 

Equations (9) and (10) are only illustrative, but already provide 
a solid argument which was 
confirmed by a set of our model calculations: (1) For given $M_{\rm wd}$, 
if $E_{\rm K}$ is larger, then $t_{+1/2}$ is smaller {\it and} $v_{\rm peak}$ 
is larger. (2) To reproduce the observationally derived value of 
$t_{+1/2}$, one can thus play with $E_{\rm K}$ (in terms of $f_{\rm ECE}$, for example). 
The model at the same time predicts smaller $v_{\rm peak}$ for smaller $M_{\rm 56Ni}$. 

The combination ($t_{+1/2}$, $v_{\rm Si II}$) of SN 2006gz is totally consistent with the 
expectation from $M_{\rm 56Ni} \sim 1 M_{\sun}$. On the other hand, 
the above relation between $t_{+1/2}$ and $v_{\rm peak}$ 
raises a difficulty in interpreting the observational data of SN 2003fg, 
since ($t_{+1/2}$, $v_{\rm Si II}$) falls into the range even below the 
lower boundary of the Chandrasekhar model with $M_{\rm 56Ni} = 0.6 M_{\sun}$. 

\subsection{Peak Luminosity and Phillips Relation}

\begin{figure}
\includegraphics[width=84mm]{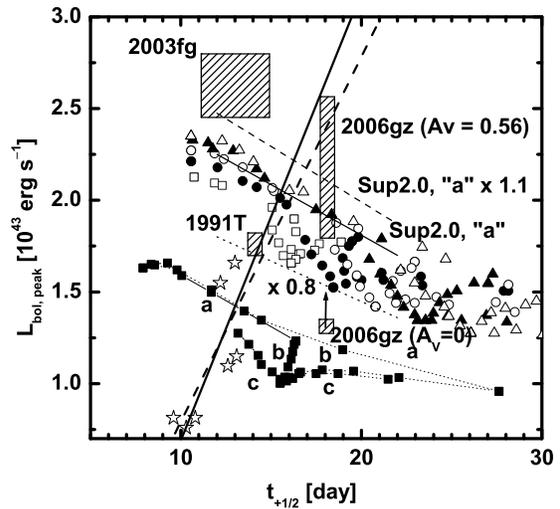}
\caption{$L_{\rm bol, peak}$ as a function of $t_{+1/2}$. 
See the caption of Fig. 3 for the meaning of the symbols,  
For SN 2006gz, shown here are the range of the luminosity 
from Hicken et al. (2007) with likely $A_{V} = 0.56$ for 
the host extinction, and the stringent lower limit with $A_{V} = 0$ 
(see Maeda et al. 2008b). 
SN Ia samples from Contardo et al. (2000) are shown by open stars 
(among which SN Ia 1991T is explicitly indicated, but with the luminosity 
from Stritzinger et al. 2006). 
The observationally-derived relation between the 
$t_{+1/2}$ and $L_{\rm bol, peak}$ 
(equivalent to the Phillips relation) is shown by a solid line 
(excluding bright SN Ia 1991T) 
and by a dashed line (excluding bright SN Ia 1991T and faint SN Ia 1991bg). 
An approximate linear fit to the results of the Sup2.0 sequence 
(with mixing "a") is indicated by the solid line. 
Two lines ("x1.1" shown by dashed line, "x0.8" shown by dotted line) roughly indicate 
what is expected for models with $M_{\rm 56Ni} = 1.1 M_{\sun}$ and $0.8 M_{\sun}$, 
respectively. 
}
\label{fig5}
\end{figure}

The $t_{+1/2} - v_{\rm peak}$ relation is the main point in the 
present paper. We also examine the bolometric peak luminosity 
($L_{\rm bol, peak}$) for the self-consistency test in this section.

Figure 5 shows $L_{\rm bol, peak}$ as a function of $t_{+1/2}$ 
for all the models (except for Sup2.0IM10/CO10). 
Normal SNe Ia (shown in stars in Fig. 5) can basically explained by 
the SW7 model sequence, if we allow dispersion in $M_{\rm 56Ni}$ 
(i.e., $L_{\rm peak} \propto M_{\rm 56Ni}$ if other properties 
are the same). 
Our model value ($M_{\rm 56Ni} \sim 1 M_{\sun}$) is slightly 
smaller than required by the peak luminosity of SNe 2003fg. 
The best value is $M_{\rm 56Ni} \sim 1.2 M_{\sun}$, 
as is consistent with the estimate by Howell et al. (2006). 
For SN 2006gz, the best value is $M_{\rm 56Ni} \sim 1.1 M_{\sun}$, 
if we adopt $E(B- V)_{\rm host} = 0.18$ and $R_{V} = 3.1$ 
(Hicken et al. 2007). The stringent lower limit for SN 2006gz is given by 
setting the host extinction negligible (\S 2); $M_{\rm 56Ni} \ga 0.7 M_{\sun}$. 

With these values, let us go back to the discussion on the $t_{+1/2} - v_{\rm peak}$ 
relation. If we take $M_{\rm 56Ni} = 1.1 M_{\sun}$ for SN 2006gz, 
the lower boundary in Fig. 4 moves 
slightly to the right, but still marginally consistent with SN 2006gz. 
For SN 2003fg, further increasing $M_{\rm 56Ni}$ makes the situation worse: 
the lower limit of $v_{\rm peak}$ now increases, making the deviation larger. 
Thus, the conclusion in the previous section does not change. 

So far we have seen that properties of SN 2006gz is consistent 
with super-Chandrasekhar models. Then, can we identify a set of 
model parameters ($M_{\rm wd}$ and $M_{\rm 56Ni}$) relevant to this SN? 
The $t_{+1/2} - v_{\rm Si II}$ constraint can be satisfied as long as 
$M_{\rm 56Ni} \la 1.1 M_{\sun}$. This is only the upper limit 
corresponding to the condition $v_{\rm peak} \la v_{\rm Si II}$.  
On the other hand, $L_{\rm peak}$ requires $M_{\rm 56Ni} \ga 1 M_{\sun}$ 
for $A_{V} = 0.56$ and $M_{\rm 56Ni} \ga 0.7 M_{\sun}$ for $A_{V} = 0$. 
Thus, we conclude that an explosion producing $M_{\rm 56Ni} \sim 1 - 1.1 M_{\sun}$ 
is the most likely explanation for SN 2006gz, with possible range of 
$M_{\rm 56Ni} \sim 0.7 - 1.1 M_{\sun}$. 
Directly deriving $M_{\rm wd}$, rather than $M_{\rm 56Ni}$, turns out to be difficult: 
$M_{\rm wd}$ is degenerated in the $t_{+1/2} - v_{\rm peak}$ relation. 
In terms of $L_{\rm peak}$, $M_{\rm wd}$ 
is again basically degenerated according to the approximate relation 
$L_{\rm bol, peak} \propto M_{\rm 56Ni}$. 
Although there is a diversity resulting from different $M_{\rm wd}$, 
the model luminosity differs up to only $\sim 10$ per cent 
for the same mixing prescription, 
between $M_{\rm wd} = 1.4 M_{\sun}$ and $2.6 M_{\sun}$. 
The variation is smaller than the observed error in $L_{\rm bol, peak}$ 
($\sim 20$ per cent), 
and thus it is difficult to derive $M_{\rm wd}$ from the analysis presented in this paper. 

\subsection{Temperature and Peak Luminosity}
Additional test on the models can be provided by the ejecta temperature. 
The discussion in this section is only qualitative: 
deriving the photospheric temperature requires detailed spectrum modeling 
(e.g., Hachinger et al. 2008), which is beyond the scope of the present study. 

The ratio of equivalent widths of Si II $\lambda$5972 to Si II $\lambda$6355 
provides a temperature indicator. In normal SNe Ia, 
the ratio is smaller for SNe with larger peak luminosity, and thus higher temperature 
(Nugent et al. 1995; Hachinger et al. 2008). The ratio in SN 2003fg is 
similar to that in relatively faint SN Ia 1994D (fig. 3 of Howell et al. 2006), 
and that in SN 2006gz is similar to normal SN Ia 2003du (fig. 1 of Hicken et al. 2007). 
These infer that the temperature of SN 2003fg is lower than 
normal SNe Ia with $M_{\rm 56Ni} \sim 0.6 M_{\odot}$, and that in SN 2006gz is 
comparable to the normal case. 

Figure 6 shows the effective temperature at bolometric maximum brightness 
in our models. The effective temperature is generally a decreasing function of 
$t_{+1/2}$; Although the photospheric velocity is smaller for larger $t_{+1/2}$ (Fig. 4), 
the peak luminosity is smaller and the photospheric radius (i.e., the photospheric 
velocity multiplied by the peak date) is larger for larger $t_{+1/2}$. The effect of 
the latter functions is more important, resulting in the dependence in Figure 6. 
For models resulting in same $t_{+1/2}$, the temperature is larger for larger 
$M_{\rm wd}$ (see Fig. 4: The photospheric velocity is slightly smaller for 
larger $M_{\rm wd}$). 

The temperature of SN 2003fg looks to be lower than normal ( i.e., SW7 models), 
and $t_{+1/2}$ is comparable to normal SNe Ia (i.e., $t_{+1/2} \sim 12 - 14$ days). 
This requires that the ejecta mass is smaller than in SW7. This is again consistent 
with our conclusion independently derived using the photospheric velocity. 
The temperature of SN 2006gz is comparable to normal SNe Ia, 
but $t_{+1/2}$ ($\sim 18$ days) is larger than normal cases. This indicates that 
the ejecta mass is larger than in normal SNe Ia (Fig. 6). 

\begin{figure}
\includegraphics[width=84mm]{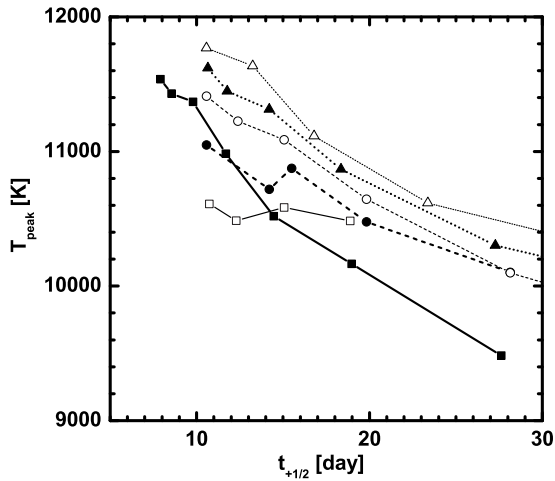}
\caption{The effective temperature at bolometric maximum brightness, 
as a function of $t_{+1/2}$. 
The meaning of the symbols are same as in Figure 4. 
Models shown here are for the mixing case "a".}
\label{fig6}
\end{figure}

\section{Discussion}

\subsection{Evaluation of Uncertainties}

\begin{figure}
\includegraphics[width=84mm]{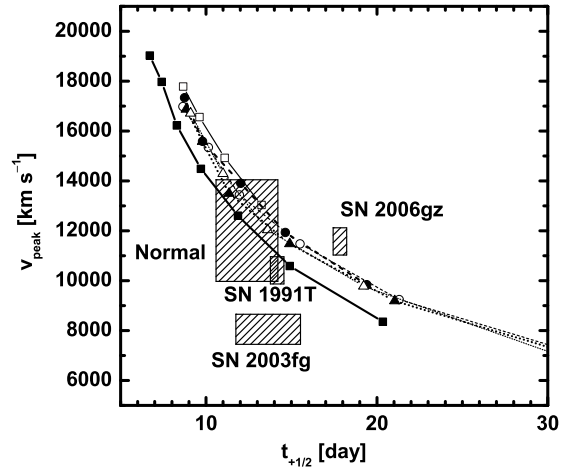}
\caption{The same as Figure 4, but for different opacity prescription, 
independent from composition. 
}
\label{fig7}
\end{figure}

One may wonder that uncertainty involved in our calculations might 
change our conclusion. In this section, we examine dominant sources of 
the uncertainty in our calculations. Here we examine effects of 
(a) different opacity prescription and (b) density distribution. 
In short, these do not alter our conclusion. 

\subsubsection{Opacity}

We have performed the same calculations for all the models, with 
different opacity prescription. Here, the opacity takes the following form: 
$\kappa_{\rm line} = 0.05$ cm$^{2}$ g$^{-1}$, independent from 
composition. The result is shown in Figure 7. 
The combination ($t_{+1/2}$, $v_{\rm peak}$) of the Super-Chandrasekhar WD model is again 
above that of the Chandrasekhar model and above ($t_{+1/2}$, $v_{\rm Si II}$) 
of SN 2003fg. The observed values of SN 2006gz are slightly above the lower boundary of the 
Super-Chandrasekhar model, and thus this SN is quite consistent with the Super-Chandrasekhar 
explosion. In this case, $v_{\rm peak} \propto M_{\rm wd}$ as the opacity is independent 
from the composition. The same conclusion 
as in the previous section applies, only if we use $M_{\rm wd}$ as the major function 
rather than $M_{\rm 56Ni}$: To explain the observational characteristics of SN 2003fg, 
we require $M_{\rm wd}$ smaller than in the normal SNe Ia.

\subsubsection{Density Distribution}

\begin{figure}
\includegraphics[width=84mm]{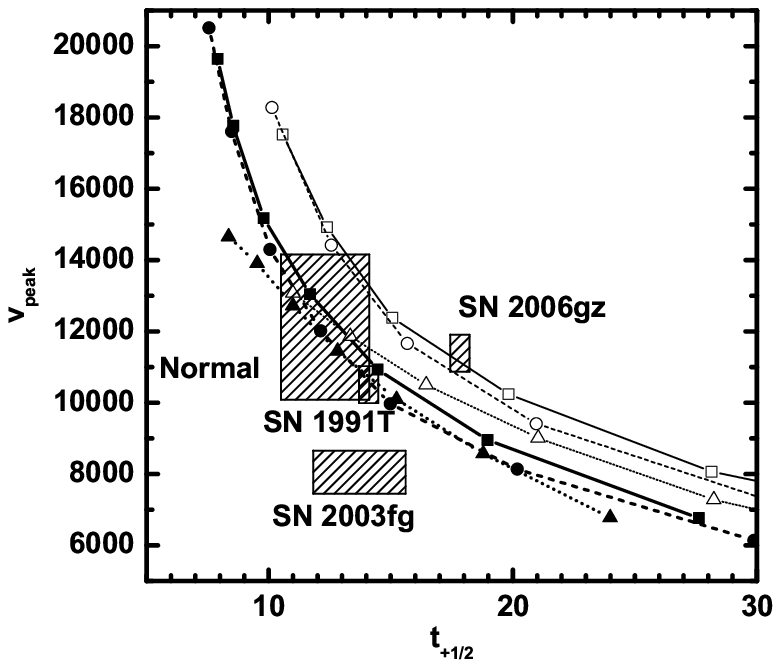}
\caption{The same as Figure 4, but for models with exponential or constant density 
distribution. Shown here are the following: 
exponential distribution with $M_{\rm wd} = 1.39 M_{\sun}$ and $M_{\rm 56Ni} = 0.6 M_{\sun}$ 
(filled circles and thick dashed line), 
exponential with $M_{\rm wd} = 2 M_{\sun}$ and $M_{\rm 56Ni} = 1 M_{\sun}$ 
(open circles and thin dashed line), 
constant density with $M_{\rm wd} = 1.39 M_{\sun}$ and $M_{\rm 56Ni} = 0.6 M_{\sun}$ 
(filled triangles and thick dotted line), 
constant with $M_{\rm wd} = 2 M_{\sun}$ and $M_{\rm 56Ni} = 1 M_{\sun}$ 
(open triangles and thin dotted line). For comparison, 
the W7-sequence is also shown (SW7CO by filled squares and thick solid, 
Sup2.0CO by open squares and thin solid).
} 
\label{fig8}
\end{figure}

The models with exponential or constant density distributions are examined 
for selected model parameters to check the uncertainty. 
Most of existing explosion models (including the W7 model) predict more or less exponential 
density distribution (e.g., Woosley et al. 2007). 
We constructed the exponential density distribution from $M_{\rm WD}$ and 
$E_{\rm K}$ according to description in Jeffery et al. (2006) and Woosley et al. (2007). 
The construction of the constant density distribution is trivial. 
These two types of density distribution are examined only for the following cases: 

\begin{enumerate}
\item $M_{\rm wd} = 1.39 M_{\sun}$ and $2 M_{\sun}$
\item $\rho_{\rm c} = 3 \times 10^9$ g cm$^{-3}$
\item $f_{\rm 56Ni} = 0.43$ ($M_{\rm wd} = 1.39 M_{\sun}$) or $0.50$ ($M_{\rm wd} = 2 M_{\sun}$)
\item $f_{\rm IME} = 0$
\item mixing $=$ "a" 
\end{enumerate}
For each model, $f_{\rm ECE}$ is varied to form the lower boundary in 
the $v_{+1/2} - v_{\rm peak}$ plot for given $M_{\rm wd}$. 

The result is shown in Figure 8. Exponential models are similar to the W7 density models 
in the $t_{+1/2} - v_{\rm peak}$ plane. The constant density distribution predicts 
smaller $v_{\rm peak}$ than in the other models, but not drastically change the result. 
The constant density distribution is an extreme assumption, probably resulting 
in the lowest value of $v_{\rm peak}$ for given $t_{+1/2}$. Thus, the uncertainty in the 
density distribution, as long as spherical symmetry is assumed, does not change our 
conclusion. 

\subsection{SN 2003fg}

In this section (\S 5.2) and the following two sections (\S\S 5.3 \& 5.4), 
we summarize our results for individual objects, and discuss implications. 
For SN 2003fg, we have found that the observational characteristics 
cannot be put into the super-Chandrasekhar-mass WD explosion scenario; 
Most importantly, the observed $t_{+1/2} - v_{\rm peak}$ relation (\S 4.1) requires that 
either $M_{\rm 56Ni}$ or $M_{\rm wd}$ (or both) should be smaller than 
even the Chandrasekhar mass, contrary to the earlier expectations (Howell et al. 2006). 
Additional support is provided by the ejecta temperature (\S 4.3), which also 
indicates that the ejecta mass (i.e., $M_{\rm wd}$) is smaller than the 
Chandrasekhar mass. On the other hand, the large peak luminosity requires 
that $M_{\rm WD} \sim 1.1 M_{\odot}$ (\S 4.2; Howell et al. 2006). 

This is apparently a contradiction. 
To remedy the problem, we suggest that the ejecta 
structure is far from spherical. 
Our radiation transfer calculations in this paper 
assume spherical symmetry, and for example 
proportional coefficient in equation (8) for $v_{\rm peak}$ should be 
a function of the viewing angle in the presence of 
large deviation from spherical symmetry. If the viewing angle is such that 
the effective (isotropic) mass is small along the line-of-sight, 
this may effectively look like an explosion with small amount of $^{56}$Ni and/or 
$M_{\rm wd}$, although the large luminosity can probably be provided by $^{56}$Ni 
in the whole ejecta, not only along the light-of-sight (note that the photospheric 
velocity is more sensitively affected than the peak luminosity in asymmetric 
SN models; Maeda et al. 2006b; Tanaka et al. 2007). 

The important finding here is that the mass should be effectively small 
as viewed from an observer to satisfy the $t_{+1/2} - v_{\rm peak}$ constraint. 
Such an explosion can not be a strongly jetted explosion of a spherically symmetric 
progenitor star: 
The jet-type explosion should yield smaller $t_{+1/2}$ for an observer closer to the jet 
direction (thanks to the large isotropic $E_{\rm K}/M_{\rm WD}$) but at the same time 
resulting in larger $v_{\rm peak}$ (for the same reason) (see e.g., 
Maeda et al. 2006b; but see Hillebrandt, Sim, \& R\"opke 2007). 
Alternatively, we suggest that {\it the progenitor star} is highly 
aspherical. This may actually be consistent with the Super-Chandrasekhar model, 
as such a massive WD should rotate rapidly to support an exceeding mass. 

Although an explosion based on such a deformed configuration has not been examined 
except for the purely detonation model (Steinmetz, M\"uller, \& Hillebrandt 1992), 
we believe it is rational to assume that the structure after the explosion 
preserves the initial configuration to some extent. The disc-like structure 
has effectively small isotropic mass if viewed along the axis of the 
rotational symmetry. We suggest this is a situation in SN 2003fg. 

Hillebrandt et al. (2007) suggested an off-centre explosion model 
(a kind of one-sided jet-like explosion model) within the context of a 
Chandrasekhar-mass WD explosion, as an alternative explanation for SN 2003fg 
(see also Sim et al. 2007). 
Their argument is based on the viewing angle effect on the light curve features, 
i.e., the larger luminosity and smaller diffusion time-scale for 
an observer closer to the direction of the $^{56}$Ni-rich blob. 
We emphasize the differences between our and their works: 
(1) Our suggestion on the ejecta asymmetry is based on the combination of the 
light curve and spectral features, and (2) the disc-like/oblate geometry, 
as we favor in this work, is different from their suggestion. 

Although we have clarified the need for the ejecta asymmetry for SN 2003fg, 
the progenitor WD mass ($M_{\rm wd}$) is not conclusively constrained 
by the present study. This will need the detailed multi-dimensional radiation 
transfer calculations; especially, we need to understand 
the dependence of the luminosity on the geometry and the viewing angle.

\subsection{SN 2006gz}

We have found that observed features of 
SN 2006gz are consistent with expectations from 
the super-Chandrasekhar-mass WD explosion scenario. 
A difficulty in identifying the underlying model for SN 2006gz 
is the uncertainty of its host's extinction and the luminosity. 
Although we cannot perfectly reject the possibility that this is 
an explosion of a Chandrasekhar WD, we favor that this is indeed 
a super-Chandrasekhar WD explosion; the $t_{+1/2} - v_{\rm peak}$ 
relation (\S 4.1) and the ejecta temperature (\S 4.3) are consistent with 
the expectations from a super-Chandrasekhar WD explosion. 

For SN 2006gz, it is not necessary to introduce the ejecta asymmetry, 
unlike SN 2003fg. These two SNe may be intrinsically different, 
or, SN 2006gz may be explained by the configuration similar to that of SN 2003fg, 
but viewed at a different angle.  Here, 
we point out that the difference in the observational features of these two SNe are qualitatively 
consistent with expected difference arising from the same configuration 
but viewed at different orientations. First, 
these two SNe Ia require similar amount of $^{56}$Ni as estimated from the peak luminosity, 
and the main difference is in 
the spectroscopic feature ($v_{\rm peak}$) and the light curve width ($t_{+1/2}$). 
The latter two are expected to be 
more sensitively dependent on the viewing angle than the estimated value of $M$($^{56}$Ni) 
(Maeda et al. 2006b). 
For the disc-like/oblate configuration, we expect larger $v_{\rm peak}$ and 
larger $t_{+1/2}$ for the larger inclination. These two values in SNe 2003fg 
and 2006gz follow this tendency. 
This is, however, only a qualitative argument, and discriminating these two possibilities 
(the difference in the ejecta shape or the viewing angle) 
needs more detailed, multi-dimensional study.

Maeda et al. (2008b) reported that SN 2006gz is peculiar also at late phases, 
$\ga 300$ days after the explosion. Applying the standard $^{56}$Ni/Co/Fe 
heating scenario, they estimated $M_{\rm 56Ni}$ by a factor of 5 
smaller that estimated with the early-phase data\footnote{This is 
independent from the uncertainty in the host extinction.}. 
This contradiction could mean either of the following two possibilities; 
(1) SN 2006gz was not powered by decays of $^{56}$Ni/Co/Fe at the early-phases 
(which then casts doubt on the super-Chandrasekhar-mass WD progenitor), 
or (2) it was powered by the decay, but for some reason (e.g., thermal catastrophe 
or dust formation), the bulk of the emission might be shifted to NIR/Mid-IR. 
Based on the result in this paper, we favor the second possibility. 

\subsection{SN 1991T}
For SN 1991T, $t_{+1/2}$ is at the upper boundary of normal SNe Ia, and  $v_{\rm Si II}$ 
is at the lower boundary. This is marginally explained by the SW7 model sequence 
with $M_{\rm 56Ni} = 0.6 M_{\sun}$. 
Thus, this constraint results in $M_{\rm 56Ni} \la 0.6 M_{\sun}$. 
On the other hand, the other constraint from $L_{\rm bol, peak}$ 
results in $M_{\rm 56Ni} \ga 0.8 M_{\sun}$ (Fig. 5). 

Strictly speaking, these two constraints are not mutually consistent, 
and the same argument for SN 2003fg can apply to SN 1991T.  
However, the deviation between the observations and models is not 
as large as in SN 2003fg, and SN 1991T may be marginally 
consistent with a spherical explosion with 
$M_{\rm 56Ni} \sim 0.8 M_{\sun}$ within uncertainties involved in our 
model calculations. 

Alternatively, SN 1991T may also be an aspherical, disc-like explosion 
like our suggestion for SN 2003fg. However, the intrinsic property 
(e.g., the explosion geometry, the mass of $^{56}$Ni) of SN 1991T 
is likely different from that of SN 2003fg (note that there is a 
possibility that SN 2006gz is intrinsically similar to SN 2003fg). 
The estimated value of $M$($^{56}$Ni) is different, 
and other features ($v_{\rm peak}$, $t_{+1/2}$) 
do not seem to be consistent with the expectation from the disc-like/oblate 
geometry which we favor for SN 2003fg.

\section{Concluding Remarks}

In this paper, we critically examined super-Chandrasekhar-mass WD 
models for (candidate) over-luminous SNe Ia 2003fg, 2006gz, and 
moderately over-luminous SN Ia 1991T. Our new approach is 
to use two observed features, $t_{+1/2}$ and $v_{\rm peak}$. 
This is equivalent to make use of a combined information of the 
light curve and spectra, and thus is more powerful than previous studies 
which mainly relied on the peak luminosity. 
Our conclusions are summarized as follows: 
\begin{enumerate}
\item Somewhat negatively, the observations of SN 2003fg are not readily 
explained by the standard $^{56}$Ni/Co/Fe heating scenario. 
The combination of relatively small $t_{+1/2}$ and small $v_{\rm peak}$ requires that 
either $M_{\rm wd}$ or $M_{\rm 56Ni}$ (or both) should be smaller than in 
normal SNe Ia. 
\item The observations of SN 2006gz can be naturally accounted for, 
by the (spherical) super-Chandrasekhar-mass model with $M_{\rm 56Ni} \sim 1 - 1.1 M_{\sun}$. 
Although the peak luminosity of SN 2006gz is largely uncertain (\S 2), 
we found that the super-Chandrasekhar-mass model is consistent with 
various observational features of SN 2006gz, and thus we favor the interpretation 
that SN 2006gz was indeed over-luminous at peak as suggested by 
Hicken et al. (2007). 
\item Observed features of SN 1991T is marginally explained by 
a spherical explosion of a WD and $M$($^{56}$Ni) $\sim 0.8 M_{\odot}$. 
This may either be a Chandrasekhar WD or a super-Chandrasekhar explosion. 
\end{enumerate}

The failure of fitting SN 2003fg is {\it not} a result of the $M_{\rm wd} - E_{\rm K}$ 
relation expected for SNe Ia. In short, 
the large amount of $^{56}$Ni should inevitably yield too large diffusion time-scale. 
One can then try a model with large $E_{\rm K}$ to reduce the diffusion time-scale, 
but this inevitably leads to the large velocity - thus the contradiction. 
Therefore, the observed characteristics of SN 2003fg are indeed inconsistent with 
any parameter set within the standard $^{56}$Ni/Co/Fe heating scenario. 

We suggest a solution to remedy 
the problem - the ejecta asymmetry resulting from a disc-like or oblate 
density structure of a rapidly rotating progenitor WD. 
This may indeed be consistent with the super-Chandrasekhar WD scenario.  
Within the results obtained by the present calculations, however, 
we can not conclusively derive the progenitor WD mass. 
An interesting possibility is that 
the different observational features of 
SNe 2003fg and 2006gz may be unified into one scheme, i.e., 
an explosion of super-Chandrasekhar WD viewed at different directions. 

Thomas et al. (2002) pointed out that a signature of 
ejecta asymmetry can be identified in the absorption strength of Si II $\lambda$6355. 
They concluded that the typical scale of inhomogeneity should be 
smaller than the size of the photosphere in normal SNe Ia, 
in order to account for uniformity of the absorption depth of the Si II. 
From fig. 1 of Hicken et al. (2007), we see that the absorption strength is 
indeed different between SNe 2003fg and 2006gz; SN 2003fg seems to show the 
weaker absorption. This may support the hypothetical configuration 
mentioned above, i.e., disc-like ejecta structure with 
SN 2003fg viewed closer to the polar than SN 2006gz; In the polar direction, 
there are a small amount of materials to absorb the light emitted at the photosphere, 
leading possibly to the weaker absorption strength in SN 2003fg.

Our results can be tested by future observations of over-luminous SNe Ia. 
First, we expect a large degree of polarization in SN 2003fg-like over-luminous 
SNe Ia, depending on the viewing direction. If SN 2006gz is an explosion similar 
to SN 2003fg, but viewed at different orientation, we also expect a large 
degree of polarization for SN 2006gz-like SNe Ia. 
Next, late-time nebular spectra may be used to infer the 
distribution of $^{56}$Ni directly\footnote{However, according to 
late-time observations of SN 2006gz,  
it did not show strong [Fe II] to probe the distribution of $^{56}$Ni at $\sim 1$ year 
after the explosion (Maeda et al. 2008b). 
Some unidentified emission process at the late-phase 
might cause this unexpected behavior (\S 5.3), and spectroscopy at relatively early epochs is 
therefore recommended for the study of geometry for these peculiar SNe Ia. 
} (Motohara et al. 2006; Maeda et al. 2008a; Modjaz et al. 2008). 
Furthermore, if we plot a number of over-luminous SNe Ia on the $t_{+1/2} - v_{\rm peak}$ 
plot, it will tell in a statistical way how large the deviation from spherical symmetry 
on average is, and how much the intrinsic diversity (e.g., $M_{\rm wd}$) in the 
super-Chandrasekhar-mass WD explosions is. 

More detailed, multi-dimensional studies are 
necessary to confirm our suggestions and speculations. First, multi-dimensional hydrodynamic 
simulations, based on a rapidly rotating super-Chandrasekhar WD, 
should be useful to check if our preferred geometry, i.e., the disc-like or oblate 
distribution of density and $^{56}$Ni, can indeed be realized. Next, multi-D 
radiation transport calculations should be 
useful to see if the $t_{+1/2} - v_{\rm peak}$ constraint 
can indeed be satisfied. We postpone these studies to the future, since in this paper 
we concentrate on examining a large parameter space to clarify the applicability and 
difficulty in spherically symmetric models (practically impossible to do in expensive 
multi-dimensional study), and to clarify the need for the non-spherical ejecta 
(for SN 2003fg). 

Identifying the geometry should be useful to evaluate possible contamination 
of SNe Ia, which do not follow the Phillips relation, in the cosmological study. 
SN 2006gz is roughly consistent with the Phillips relation if we adopt 
the peak luminosity derived by Hicken et al. (2007) 
(\S 4.2 and \S 4.3; Fig. 5), while SN 2003fg is not (Howell et al. 2006). 
Our scenario for the disc-like/oblate ejecta infers that over-luminous SNe Ia 
which do not follow the Phillips relation (e.g., SN 2003fg), should also show 
the peculiar low photospheric velocity. 
Thus, by performing spectroscopy, the SNe Ia with large deviation from 
the Phillips relation can be easily identified, and can be removed from 
samples for the cosmological study. 

\section*{Acknowledgments}
The authors would like to thank Ken'ichi Nomoto and 
Masaomi Tanaka for useful discussion, and the anonymous 
referee for many constructive comments. 
This research has been supported by World Premier International Research 
Center Initiative (WPI Initiative), MEXT, Japan, and by the 
Grant-in-Aid for Young Scientists of the JSPS/MEXT (20840007). 
This research has made use of the CfA Supernova Archive, 
which is founded in part by the National Science Foundation 
through grant AST 0606772.

\bsp

\label{lastpage}

\end{document}